# Anomalous scaling behavior in a mixed-state Hall effect of a cobalt-doped BaFe$_2$As$_2$ epitaxial film with a high critical current density over 1 MA/cm$^2$

Hikaru Sato[1], Takayoshi Katase[2], Won Nam Kang[1, 3], Hidenori Hiramatsu[1],

Toshio Kamiya[1], and Hideo Hosono[1, 2, 4 *]

1: *Materials and Structures Laboratory, Tokyo Institute of Technology, Mailbox R3-1, 4259 Nagatsuta-cho, Midori-ku, Yokohama 226-8503, Japan*

2: *Frontier Research Center, Tokyo Institute of Technology, S2-6F East, Mailbox S2-13, 4259 Nagatsuta-cho, Midori-ku, Yokohama 226-8503, Japan*

3: *BK21 Physics Division and Department of Physics, Sungkyunkwan University, 300 CheonCheon-dong, Jangan-gu, Suwon, Gyeonggi-do 440-746, Republic of Korea*

4: *Materials Research Center for Element Strategy, Tokyo Institute of Technology, 4259 Nagatsuta-cho, Midori-ku, Yokohama 226-8503, Japan*

## Abstract

The mixed-state Hall effect was examined in a Ba(Fe$_{1-x}$Co$_x$)$_2$As$_2$ epitaxial film with a high critical current density. The transverse resistivity $\rho_{xy}$ and the longitudinal resistivity $\rho_{xx}$ follow power law scaling $\rho_{xy} = A\rho_{xx}^{\beta}$. In the temperature-sweep with a fixed field ($T$ sweep), all of the $\beta$ values are independent of magnetic field up to 9 T, and are lower than 2.0 (around 1.8). In contrast, the $\beta$ values in the magnetic-field sweep with a fixed temperature ($H$ sweep) change from 1.8 to 2.0 as the temperature increases from 13 to 16 K even in the $T/H$ region that overlaps with the $T$ sweep measurements. These results indicate that the vortices introduced at low temperatures are trapped by strong pinning centers, but a portion of the vortices introduced at high temperatures are not strongly trapped by the pinning centers. The sign of $\rho_{xy}$ is negative, and a sign reversal is not detected. These distinct scaling behaviors, which sharply contrast cuprates and MgB$_2$, are explained by high-density $c$-axis pinning centers in the Ba(Fe$_{1-x}$Co$_x$)$_2$As$_2$ epitaxial film and are consistent with a wider vortex liquid phase.

## I. Introduction

Since the discovery of an iron-pnictide superconductor with a critical temperature ($T_c$) of 26 K,[1] $T_c$ has been improved and now reaches 55 K.[2] To investigate the intrinsic properties of iron-pnictide superconductors and to apply them to practical devices, the first thin film was epitaxially grown in 2008.[3] In particular, Ba(Fe$_{1-x}$Co$_x$)$_2$As$_2$ (BaFe$_2$As$_2$:Co) epitaxial films have achieved high critical current densities ($J_c$) $\geq 1$ MA/cm$^2$,[4-6] and have been applied to Josephson junctions[7,8] and a superconducting quantum interference device.[9]

Compared to cuprates, the properties of iron-pnictide Josephson-junction devices remain inferior due largely to their metallic characters. However, their superior grain boundary properties[10] along with high upper critical magnetic fields ($H_{c2}$) $\geq 50$ T and small anisotropic factors $\gamma = H_{c2}^{//ab} / H_{c2}^{//c} = 1-2$ [11] are promising for wire[12-14] and tape applications.[15,16] Similar to cuprates, the extrinsic pinning centers largely control $J_c$ in iron pnictides. Therefore, the vortex pinning mechanism for high-$J_c$ BaFe$_2$As$_2$:Co epitaxial films is important to understand the ultimate potential of the iron pnictide superconducting wires and tapes. Although the pinning properties have been studied extensively on single crystals,[17,18] there are few reports on high-$J_c$ pnictide films, particularly about transport measurements.[19]

Because most high-$J_c$ BaFe$_2$As$_2$:Co epitaxial films are grown on electrically conductive buffer layers such as Fe metal[20] and SrTiO$_3$,[4] it is difficult to extract reliable transport properties. On the other hand, we have successfully grown BaFe$_2$As$_2$:Co epitaxial films with $J_c > 1$ MA/cm$^2$ directly on insulating single-crystal substrates without a buffer layer,[5,21] allowing the transport properties of iron-pnictide epitaxial films to be directly and reliably examined.

The Hall effect measurement in a normal–superconducting mixed-state is a representative probing technique of a vortex motion. High-$T_c$ superconductors such as cuprates and MgB$_2$ follow a power law scaling of $\rho_{xy} = A\rho_{xx}^{\beta}$, where $\rho_{xy}$ and $\rho_{xx}$ are the transverse and longitudinal resistivities, respectively. The behavior of the $\beta$ value has actively been discussed in relation to the superconducting and vortex pinning mechanisms.[22-27]

Although sign reversal of $\rho_{xy}$ has been observed, its origin remains controversial.[28-30] Vinokur *et al.*[31] considered the momentum balance and proposed a model where a system of interacting vortices under quenched disorders and thermal noises follows a universal scaling law with a universal $\beta = 2.0$. However, this model does not explain all the mixed-state Hall effects of high-$T_c$ superconductors. Wang *et al.*[32] considered both backflow currents and thermal fluctuations, and suggested that the $\beta$ value varies from 2.0 to 1.5 as the pining strength increases. Comparison of the experimental results to theoretical models is important to understand the vortex dynamics in a mixed-state of high-$J_c$ BaFe$_2$As$_2$:Co epitaxial films.

In this study, we examine the transport properties in a mixed-state high-quality BaFe$_2$As$_2$:Co epitaxial film with a high $J_c$ over 1 MA/cm$^2$. The pinning mechanism is discussed based on the obtained $\beta$ values $\leq 2.0$ and different behaviors observed during the temperature ($T$) sweeps and magnetic-field ($H$) sweeps. Furthermore, the comprehensive Hall effect measurements confirm that a BaFe$_2$As$_2$:Co epitaxial film does not exhibit a sign reversal in the entire ranges of magnetic fields (1–9 T) and temperatures (13–16 K).

## II. Experimental

A 350-nm-thick $BaFe_2As_2$:Co epitaxial film was grown directly on an insulating $(La,Sr)(Al,Ta)O_3$ (001) single-crystal substrate by pulsed laser deposition (PLD) without a buffer layer. The second harmonic (the wavelength = 532 nm) of a neodymium-doped yttrium aluminum garnet laser[33] and a polycrystalline disk of $Ba(Fe_{0.92}Co_{0.08})_2As_2$ [5] were used as the excitation source and the PLD target, respectively. Detailed growth conditions and film quality are reported in the literature.[5,21] A cross-sectional image of the $BaFe_2As_2$:Co epitaxial film was observed by transmission electron microscopy (TEM). Thin samples for the TEM observations were prepared by a focused-ion-beam micro-sampling technique. Chemical composition was examined using a line-scan mode of energy dispersive x-ray spectroscopy (EDXS) with a spatial resolution of ca. 1 nm. The transition width $\Delta T_c$ of $\rho_{xx}$ (~1 K) and the transport $J_c$ at 4 K (2.4 $MA/cm^2$) under a self-field of the resulting film were comparable to those in previous papers.[5,21]

To examine the mixed-state Hall effect, the film was patterned into an eight-terminal Hall bar structure by photolithography and Ar ion milling. To minimize the contact resistance, a Au film was formed on the contact electrode pads using a lift-off process.[34] $\rho_{xx}$ and $\rho_{xy}$ were measured by applying a dc current at a density of 5 $kA/cm^2$ using two measurement modes: (i) a $T$ sweep mode with a fixed $H$ and (ii) a $H$ sweep mode with a fixed $T$. Cancelling the offset effects using the relation $\rho_{xy} = (\rho_{xy}^+ - \rho_{xy}^-)/2$, where $\rho_{xy}^+$ and $\rho_{xy}^-$ were measured under inverted magnetic fields, the net $\rho_{xy}$ values were extracted. The measurement limit of our system was ~125 nV due to fluctuations in $T/H$ and the digital voltmeter. Thus, the reliability was for $\rho_{xy} > 5.0 \times 10^{-9}$ $\Omega$cm and $\rho_{xx} > 1.4 \times 10^{-10}$ $\Omega$cm. Magnetic fields were applied parallel to the *c*-axis of the film and varied up to 9 T.

## III. Results

Figure 1(a) shows a cross sectional TEM image of $BaFe_2As_2$:Co film. The heterointerface between the substrate and $BaFe_2As_2$:Co is very sharp and no sub-product is observed. The crystallographic orientation between the substrate and the film is the same as that of a previously-reported our film [5,21]. In addition, some line defects like domain boundaries along the *c*-axis are observed in the film as indicated by the vertical arrows. We confirmed that the chemical composition at the defect is exactly the same as the film bulk region [Fig. 1(b)]. These lateral size are estimated to be 2–5 nm and roughly double of the coherence length of $BaFe_2As_2$:Co;[35] such sizes of defects effectively work as pinning centers, and the existence of defects along the *c*-axis is consistent with $J_c$ peak along the *c*-axis in the results of the angle dependence of $J_c$ [19,36]

Figures 2 (a) and (b) show the *T* sweep results of $\rho_{xx}$ and $\rho_{xy}$ under fixed magnetic fields varied up to 9 T, respectively. As *H* increases, $\Delta T_c$ broadens along with shifts in $T_c^{onset}$ and $T_c^{zero}$ toward lower temperatures. As discussed later, this broadening originates from the presence of a vortex liquid state.[19] The negative $\rho_{xy}$ [Fig. 2(b)] indicates that the majority of charge carriers are electrons, which is consistent with aliovalent ion doping, i.e., the substitution of $Co^{3+}$ with $Fe^{2+}$ sites. The Hall coefficient at 25 K is $-1.7 \times 10^{-3}$ $cm^3$/C, which is comparable to those reported for single crystals with the same doping level as the present PLD target disk,[37] and implies that the Co dopant concentration in the film is almost the same as that of the PLD target disk.[38]

Figures 2 (c) and (d) show the $H$ sweep results at temperatures from 13 to 16 K for $\rho_{xx}$ and $\rho_{xy}$, respectively. Due to its high $H_{c2} \geq 50$ T,[11] superconductivity remains at 13–16 K even when high magnetic fields up to 9 T are applied. Similar to observations in $MgB_2$ films, neither the $T$ nor $H$ sweep results exhibit a Hall sign reversal.[27] However, these observations sharply contrast most cuprates such as $YBa_2Cu_3O_{7-\delta}$ (YBCO),[22] $Bi_2Sr_2Ca_2Cu_3O_{10}$ (BSCCO),[30] and $HgSr_2CaCuO_{6-\delta}$,[26] in which the sign of $\rho_{xy}$ changes from positive to negative near the superconducting transition.

Figure 3 replots the $T$ sweep results of Figs. 2 (a, b) in terms of the scaling behavior between $\rho_{xy}$ and $\rho_{xx}$. Throughout the entire mixed-state region, $\rho_{xy}$ and $\rho_{xx}$ follow the relation $\rho_{xy} = A\rho_{xx}^{\beta}$ very well. The $\beta$ values are extracted from the slopes of the double logarithmic plots by a least squares fit (the resulting standard errors $\sigma_e$ are < 0.008, except for 1 T, which is $\sigma_e = 0.03$). All the $\beta$ values are similar and definitely less than 2.0; they are between 1.7–1.8 and independent of the applied magnetic field up to at least 9 T. These $\beta$ values are lower than those reported for $Ba(Fe_{1-x}Co_x)_2As_2$ single crystals with $\beta = 3.0$–3.4 for $x = 0.08$ and $\beta = 2.0 \pm 0.2$ for $x = 0.10$,[39] but are comparable to those of YBCO epitaxial films with strong pinning centers.[22] Similar $\beta$ behaviors with magnetic fields (i.e., no variation with $H$) are also observed in BSCCO single crystals[24] and $MgB_2$ films,[27] in which weak pinning centers are introduced. However, in those two films, the $\beta$ values are close to 2.0. The above behavior for a $Ba(Fe_{1-x}Co_x)_2As_2$ film, in which a small and constant $\beta$ is maintained under various magnetic fields, is generally reported for superconductors with strong pinning centers such as twinned YBCO epitaxial films,[22] irradiated YBCO single crystals,[25] and irradiated $Tl_2Ba_2CaCu_2O_8$ films.[23,40]

Figure 4 replots the $H$ sweep results of Figs. 2 (c, d) in terms of the scaling behavior between $\rho_{xy}$ and $\rho_{xx}$. Similar to the $T$ sweep results, most of the data follow the power scaling law. However, the $\beta$ values between the $T$ and the $H$ sweeps clearly differ. Unlike the $T$ sweep results (Fig. 3), the $\beta$ value increases from 1.8 to 2.0 as the $T$ increases from 13 to 16 K. According to the above discussion for Fig. 3, this result suggests that the pinning mechanism changes as $T$ increases.

Table I summarizes the scaling results of the $T$ and the $H$ sweeps. For the $T$ sweep, the scaling behavior shows low $\beta$ values (1.7–1.8), which are independent of magnetic fields at least up to 9 T. In contrast, the $\beta$ value changes from 1.8 to 2.0 as $T$ increases in the $H$ sweep. Wang *et al.*[32] proposed a unified theory based on the normal core model[41] by taking both the backflow currents and thermal fluctuations into account. According to the Wang's theory for systems with strong pinnings, the pinning effect becomes dominant and the scaling behavior changes from $\beta \sim 2.0$ to ~1.5, indicating that a reduced $\beta$ corresponds to the weakening of the pinning strength. In fact, the $\beta$ values of most cuprate superconductors exhibiting strong pinning are 1.0–1.8, which are considerably less than 2.0.[22,23,25,26]

Based on the Wang's theory,[32] our experimental results with $\beta = 1.7$–1.82 indicate that strong pinning centers effectively work for the $T$ sweep in the entire $H$ region examined in this study. The strong pinning centers are also effective for the $H$ sweep only in the low $T$ (< 14 K) region, but the pinning strength rapidly weakens as $T$ increases and approaches $T_c$. Maiorov *et al.*[19] observed a similar behavior in the angular dependence of $J_c$; i.e., strong $c$-axis pinning remains under high magnetic fields but weakens as $T$ increases.

On the other hand, the $BaFe_2As_2$:Co epitaxial film displays an anomalous scaling behavior (i.e., different $\beta$ values are observed in the $T$ and the $H$ sweeps even for the same $T$ and the $H$ conditions). Figure 5 (a) shows the vortex phase diagram of the $BaFe_2As_2$:Co epitaxial film normalized by $H_{c2}$ and $T_c$, while Fig. 5 (b) compares the vortex phase diagrams of other superconductors. The irreversibility line $H_{irr}$, [defined by $\rho(T, H) = 0.01\rho_N$, where $\rho_N$ denotes the normal state resistivity at 25K (dotted lines)] and $H_{c2}$ [defined $\rho(T, H) = 0.90\rho_N$ (dashed lines)] are obtained from Figs. 2(a) and (c). It would be better to determine $H_{c2}$ and $H_{irr}$ by thermodynamic measurements such as heat capacity measurements, but these are difficult for thin films. We confirmed that these criterions of $H_{irr}$ and $H_{c2}$ are consistent with the values determined by $I–V$ and magnetic susceptibility measurements [38] with the criterion of $10^3$ $A/cm^2$. The horizontal and vertical lines indicate the $T$ sweep and the $H$ sweep trajectories, respectively. The corresponding $\beta$ values are indicated on the left (for $H$ sweep) and right (for $T$ sweep). The two $\beta(T, H)$ values obtained by the $T$ sweep and the $H$ sweep are similar, 1.7–1.8 in the relatively low $T$ and high $H$ region, as indicated by the dashed square area. By contrast, the two $\beta(T, H)$ values clearly differ between the $T$ sweep and the $H$ sweep even at the same $T$ and $H$ in the high $T$ and low $H$ region, as indicated by the solid square area. This behavior has not been observed in other superconductors. Because this difference is observed at the same $T$ and $H$ and depends on the sweep history, it originates from a hysteresis phenomenon probably due to the dynamics of the vortices.

We also like to note that the normalized vortex phase diagram of the $BaFe_2As_2$:Co film has a wider vortex liquid region than those of YBCO and $MgB_2$. Although the $H_{c2}$ lines are similar for all three, the $H_{irr}$ line of the $BaFe_2As_2$ film extends to the lower $T$ (i.e., to lower $H$) region [Fig. 5 (b)]. The width of the vortex liquid phase is narrower for a BaFe$_2$As$_2$:Co single crystal, indicating that the vortex liquid phase in the BaFe$_2$As$_2$:Co film is wider due to the extrinsic disorder introduced during the thin-film growth process.

## IV. Discussion

This phenomenon where the $\beta$ values depend on the measurement history shows that the vortex penetration process affects the vortex dynamics in BaFe$_2$As$_2$:Co epitaxial films, especially at a relatively high $T$ and a low $H$. Figure 6 illustrates our proposed vortices penetration model. We start with a low $T$ and zero $H$ condition in Fig. 6 (a). $U$ denotes the effective pinning potential. $N_v$ and $N_p$ denote the numbers of penetrated vortices and strong pinning centers, respectively. For the $T$ sweep, $H$ is initially applied at a low $T$, and then $T$ is swept to higher values. For the $H$ sweep, $T$ is initially adjusted while maintaining the zero $H$, and then $H$ is swept to higher values. First, we considered the case of high $H$ (i.e., the final $N_v \gg N_p$) and a relatively low $T$ (b), which corresponds to the dashed squares in Fig. 5 (a). The short and long vertical arrows denote pinned vortices and unpinned vortices, respectively. For the $T$ sweep, when a high $H$ (i.e., $N_v \gg N_p$) is applied at a low $T$ (upper left), a large portion of the vortices spills over from the pinning centers. The pinned vortices are maintained at a high $T$, while the vortices unpinned in the strong pinning centers are mobile (upper right). For the $H$ sweep, the effective pinning force weakens at higher $T$ (lower left). Then vortices are introduced at the high $T$, where most of the pinning centers are filled due to the condition $N_v \gg N_p$, and the extra vortices that spill over from the pinning centers are mobile (lower right). Thus, the difference in the $T$ and $H$ sweep data is small. The condition of low $H$ (the final $N_v \sim N_p$) and high $T$, which corresponds to the solid square area in Fig. 5 (a), provides a different result [Fig. 6(c)]. For the $T$ sweep, the pinning centers trap the penetrated vortices upon applying $H$. All the penetrated vortices are effectively trapped at the pinning centers at low $T$ because the effective pinning force is sufficiently strong (upper left), and their positions are stabilized and maintained even at a high $T$ (upper left). Consequently, a strong pinning condition with $\beta < 2.0$ is produced. By contrast, for the $H$ sweep, the effective pinning force weakens at high $T$ (lower left), and vortices are penetrated at the high $T$ with a weak pinning force. This leads to a large thermal fluctuation at high temperatures near $T_c$. Consequently, fewer vortices are trapped by the pinning centers, and the vortices are mobile (lower right), giving rise to a weak pinning condition with $\beta \sim 2.0$.

This model explains well the observation that the $\beta(T, H)$ value at the same $T$ and $H$ condition depends on the measurement history for a $BaFe_2As_2$:Co epitaxial film, e.g., the $T$ sweep or the $H$ sweep. This phenomenon may occur in other superconductors, but has only been observed in a $BaFe_2As_2$:Co epitaxial film. The above model indicates that the difference of the $\beta$ values between in the $T$ and $H$ sweep data occurs for the condition $N_v < N_p$, which suggests that the $BaFe_2As_2$:Co epitaxial film has high-density pinning centers. In $MgB_2$ and YBCO films without artificial pinning centers, it is estimated that their dominant intrinsic pinning centers are planar defects (i.e., grain boundaries and twin boundaries) due to the scaling behaviors of their normalized pinning force and transmission electron microscopy image.[42-45] On the other hand, the scaling behaviors of normalized pinning force and magnetic Bitter decoration indicate that $BaFe_2As_2$:Co has intrinsic pinning centers[11,46] and extrinsic $c$-axis pinning centers, which are also observed in Fig. 1 clearly, exist in $BaFe_2As_2$:Co epitaxial film.[19] The $F_p$ maximum position of $BaFe_2As_2$:Co epitaxial film is about 0.4 $H/H_{irr}$ [16], indicating that origin of the pinning is not purely the collective pinning and includes the strong core pining. This mechanism is similar to the case of columnar $BaFeO_2$ pinning centers [47] in a $BaFe_2As_2$:Co epitaxial film grown using $SrTiO_3$ buffer layer. These experimental results indicate the existence of a dense vortex-pinning structure in a $BaFe_2As_2$:Co epitaxial film; these conditions give a larger $N_p$ than those of other superconductors. Furthermore, because the width of the vortex liquid phase is larger in a $BaFe_2As_2$:Co epitaxial film than that of the single crystal, the $BaFe_2As_2$:Co epitaxial film has additional extrinsic pinning centers besides the intrinsic ones in the single crystal. Thus, the distinct scaling behaviors in the $BaFe_2As_2$:Co epitaxial film at a relatively high $T$ and in a low $H$ should originate from the high-density pinning structure, which consists of the intrinsic and the extrinsic pinning centers.

Unlike cuprate superconductors exhibiting $1.0 < \beta < 2.0$,[22,23,25,26] the sign of $\rho_{xy}$ in this study does not reverse in either the $T$ or the $H$ sweep results. Wang's theory[32] predicts that the scaling behavior exhibiting $\beta < 2.0$ occurs after a sign reversal, which is inconsistent with the present results. Therefore, we would like to discuss the sign reversal based on the Hall conductivity, which is defined as $\sigma_{xy} = \rho_{xy}/(\rho_{xy}^2 + \rho_{xx}^2)$, because a general argument of vortex dynamics indicates $\sigma_{xy}$ is insensitive to disorder effects.[31,40] Indeed, for cuprates, the sign reversal is clearly observed in $\sigma_{xy}$.[48,49] Figure 7 replots Fig. 2 in terms of $\sigma_{xy}$. The $H$ and the $T$ sweep results substantiate that $\sigma_{xy}$ tends to diverge to a large negative value and sign reversal is not detected. Matsuda *et al.*[48] and Nagaoka *et al.*[49] have experimentally suggested that the sign of $\rho_{xy}$ depends on the doping level in most cuprate superconductors; i.e., a sign reversal occurs in an underdoped regime but diminishes in overdoped ones. Kopnin *et al.*[50] and Dorsey[51] have also discussed the Hall anomaly based on a microscopic approach and proposed theories of a mixed-state Hall effect based on a time-dependent Ginzburg-Landau model; they postulate that the sign reversal depends on the gradient of the density of states at the Fermi surface. According to these experimental and theoretical results, researching the doping concentration dependence may provide a definitive conclusion on the Hall sign reversal in high-$J_c$ BaFe$_2$As$_2$:Co epitaxial films because this study employed a slightly overdoped sample.[21]

## V. Conclusion

We investigated the transport properties in a mixed-state of a high-$J_c$ BaFe$_2$As$_2$:Co epitaxial film. The scaling behavior in the $T$ sweep measurements shows constant $\beta$ values less than 2.0 under magnetic fields up to 9 T. On the other hand, the $\beta$ values clearly increase from 1.8 to 2.0 in the $H$ sweep measurements as $T$ increases from 13 to 16 K. These results indicate that strong pinning centers trap the vortices introduced at low temperatures, but some of the vortices introduced at high temperatures are not trapped and that pinning weakens at higher temperatures near the normal state. In the entire $H$ and $T$ ranges examined in this study, the sign does not reverse. These distinct scaling behaviors, which sharply contrast cuprates and MgB$_2$, can be explained by the high-density $c$-axis pinning centers in a BaFe$_2$As$_2$:Co epitaxial film, which are consistent with wider vortex liquid phase.

## Acknowledgment

This work was supported by the Japan Society for the Promotion of Science (JSPS) through the "Funding Program for World-Leading Innovative R&D on Science and Technology (FIRST Program). W.N.K. was partially supported by Mid-career

Researcher Program through an NRF grant funded by the Ministry of Education, Science & Technology (MEST) (No. 2010-0029136) of Korea.

(*) Corresponding author: hosono@msl.titech.ac.jp

## References

[1.] Y. Kamihara, T. Watanabe, M. Hirano, and H. Hosono, J. Am. Chem. Soc. **130**, 3296 (2008).

[2] R. Zhi-An, L. Wei, Y. Jie, Y. Wei, S. Xiao-Li, Zheng-Cai, C. Guang-Can, D. Xiao-Li, S. Li-Ling, Z. Fang, and Z. Zhong-Xian, Chin. Phys. Lett. **25**, 2215 (2008).

[3] H. Hiramatsu, T. Katase, T. Kamiya, and H. Hosono, J. Phys. Soc. Jpn. **81**, 011011 (2012).

[4] S. Lee, J. Jiang, Y. Zhang, C.W. Bark, J.D. Weiss, C. Tarantini, C.T. Nelson, H.W. Jang, C.M. Folkman, S.H. Baek, A. Polyanskii, D. Abraimov, A. Yamamoto, J.W. Park, X.Q. Pan, E.E. Hellstrom, D.C. Larbalestier, and C.B. Eom, Nat. Mater. **9**, 397 (2010).

[5] T. Katase, H. Hiramatsu, T. Kamiya, and H. Hosono, Appl. Phys. Express **3**, 063101 (2010).

[6] D. Rall, K. Il'in, K. Iida, S. Haindl, F. Kurth, T. Thersleff, L. Schultz, B. Holzapfel, and M. Siegel, Phys. Rev. B **83**, 134514 (2011).

[7] T. Katase, Y. Ishimaru, A. Tsukamoto, H. Hiramatsu, T. Kamiya, K. Tanabe, and H. Hosono, Appl. Phys. Lett. **96**, 142507 (2010).

[8] S. Schmidt, S. Döring, F. Schmidl, V. Grosse, P. Seidel, K. Iida, F. Kurth, S. Haindl, I. Mönch, and B. Holzapfel, Appl. Phys. Lett. **97**, 172504 (2010).

[9] T. Katase, Y. Ishimaru, A. Tsukamoto, H. Hiramatsu, T. Kamiya, K. Tanabe, and H. Hosono, Supercond. Sci. Technol. **23**, 082001 (2010).

[10] T. Katase, Y. Ishimaru, A. Tsukamoto, H. Hiramatsu, T. Kamiya, K. Tanabe, and H. Hosono, Nat. Commun. **2**, 409 (2011).

[11] A. Yamamoto, J. Jaroszynski, C. Tarantini, L. Balicas, J. Jiang, A. Gurevich, D.C. Larbalestier, R. Jin, A.S. Sefat, M.A. McGuire, B.C. Sales, D.K. Christen, and D. Mandrus, Appl. Phys. Lett. **94**, 062511 (2009).

[12] K. Togano, A. Matsumoto, and H. Kumakura, Appl. Phys. Express **4**, 043101 (2011).

[13] J.D. Weiss, C. Tarantini, J. Jiang, F. Kametani, A.A. Polyanskii, D.C. Larbalestier, and E.E. Hellstrom, Nat. Mater. **11**, 682 (2012).

[14] Y. Ma, Supercond. Sci. Technol. **25**, 113001 (2012).

[15] K. Iida, J. Hänisch, S. Trommler, V. Matias, S. Haindl, F. Kurth, I.L. del Pozo, R. Hühne, M. Kidszun, L. Engelmann, L. Schultz, and B. Holzapfel, Appl. Phys. Express **4**, 013103 (2011).

[16] T. Katase, H. Hiramatsu, V. Matias, C. Sheehan, Y. Ishimaru, T. Kamiya, K. Tanabe, and H.

Hosono, Appl. Phys. Lett. **98**, 242510 (2011).

[17] R. Prozorov, N. Ni, M.A. Tanatar, V.G. Kogan, R.T. Gordon, C. Martin, E.C. Blomberg, P. Prommapan, J.Q. Yan, S.L. Bud'ko, and P.C. Canfield, Phys. Rev. B **78**, 224506 (2008).

[18] Y. Nakajima, Y. Tsuchiya, T. Taen, T. Tamegai, S. Okayasu, and M. Sasase, Phys. Rev. B **80**, 012510 (2009).

[19] B. Maiorov, T. Katase, S.A. Baily, H. Hiramatsu, T.G. Holesinger, H. Hosono, and L. Civale, Supercond. Sci. Technol. **24**, 055007 (2011).

[20] T. Thersleff, K. Iida, S. Haindl, M. Kidszun, D. Pohl, A. Hartmann, F. Kurth, J. Hänisch, R. Hühne, B. Rellinghaus, L. Schultz, and B. Holzapfel, Appl. Phys. Lett. **97**, 022506 (2010).

[21] T. Katase, H. Hiramatsu, T. Kamiya, and H. Hosono, Supercond. Sci. Technol. **25**, 084015 (2012).

[22] J. Luo, T.P. Orlando, J.M. Graybeal, X.D. Wu, and R. Muenchausen, Phys. Rev. Lett. **68**, 690 (1992).

[23] R.C. Budhani, S.H. Liou, and Z.X. Cai, Phys. Rev. Lett. **71**, 621 (1993).

[24] A.V. Samoilov, Phys. Rev. Lett. **71**, 617 (1993).

[25] W.N. Kang, D.H. Kim, S.Y. Shim, J.H. Park, T.S. Hahn, S.S. Choi, W.C. Lee, J.D. Hettinger, K.E. Gray, and B. Glagola, Phys. Rev. Lett. **76**, 2993 (1996).

[26] W.N. Kang, B.W. Kang, Q.Y. Chen, J.Z. Wu, S.H. Yun, A. Gapud, J.Z. Qu, W.K. Chu, D.K. Christen, R. Kerchner, and C.W. Chu, Phys. Rev. B **59**, R9031 (1999).

[27] W.N. Kang, H.-J. Kim, E.-M. Choi, H.J. Kim, K.H.P. Kim, and S.-I. Lee, Phys. Rev. B **65**, 184520 (2002).

[28] M. Galffy and E. Zirngiebl, Solid State Commun. **68**, 929 (1988).

[29] S.J. Hagen, C.J. Lobb, R.L. Greene, M.G. Forrester, and J.H. Kang, Phys. Rev. B **41**, 11630 (1990).

[30] Y. Iye, S. Nakamura, and T. Tamegai, Physica C **159**, 616 (1989).

[31] V.M. Vinokur, V.B. Geshkenbein, M.V. Feigel'man, and G. Blatter, Phys. Rev. Lett. **71**, 1242 (1993).

[32] Z.D. Wang, J. Dong, and C.S. Ting, Phys. Rev. Lett. **72**, 3875 (1994).

[33] H. Hiramatsu, T. Katase, T. Kamiya, M. Hirano, and H. Hosono, Appl. Phys. Express **1**, 101702 (2008).

[34] W.N. Kang, H.-J. Kim, E.-M. Choi, H.J. Kim, K.H.P. Kim, H.S. Lee, and S.-I. Lee, Phys. Rev. B **65**, 134508 (2002).

[35] M. Kano, Y. Kohama, D. Graf, F. Balakirev, A.S. Sefat, M.A. Mcguire, B.C. Sales, D. Mandrus, and S.W. Tozer, J. Phys. Soc. Jpn. **78**, 084719 (2009).

[36] S.R. Foltyn, L. Civale, J.L. MacManus-Driscoll, Q.X. Jia, B. Maiorov, H. Wang, and M. Maley, Nat. Mater. **6**, 631 (2007).

[37] L. Fang, H. Luo, P. Cheng, Z. Wang, Y. Jia, G. Mu, B. Shen, I.I. Mazin, L. Shan, C. Ren, and H.-H. Wen, Phys. Rev. B **80**, 140508 (2009).

[38] H. Hiramatsu, T. Katase, Y. Ishimaru, A. Tsukamoto, T. Kamiya, K. Tanabe, and H. Hosono, Mater. Sci. Eng. B **177**, 515 (2012).

[39] L.M. Wang, U.-C. Sou, H.C. Yang, L.J. Chang, C.-M. Cheng, K.-D. Tsuei, Y. Su, T. Wolf, and P. Adelmann, Phys. Rev. B **83**, 134506 (2011).

[40] A.V. Samoilov, A. Legris, F. Rullier-Albenque, P. Lejay, S. Bouffard, Z.G. Ivanov, and L.-G. Johansson, Phys. Rev. Lett. **74**, 2351 (1995).

[41] J. Bardeen and M.J. Stephen, Phys. Rev. **140**, A1197 (1965).

[42] D. Dew-Hughes, Philos. Mag. **30**, 293 (1974).

[43] D.C. Larbalestier, L.D. Cooley, M.O. Rikel, A.A. Polyanskii, J. Jiang, S. Patnaik, X.Y. Cai, D.M. Feldmann, A. Gurevich, A.A. Squitieri, M.T. Naus, C.B. Eom, E.E. Hellstrom, R.J. Cava, K.A. Regan, N. Rogado, M.A. Hayward, T. He, J.S. Slusky, P. Khalifah, K. Inumaru, and M. Haas, Nature **410**, 186 (2001).

[44] B. Roas, L. Schultz, and G. Saemann-Ischenko, Phys. Rev. Lett. **64**, 479 (1990).

[45] T. Nishizaki, T. Aomine, I. Fujii, K. Yamamoto, S. Yoshii, T. Terashima, and Y. Bando, Physica C **181**, 223 (1991).

[46] S. Demirdiş, C.J. van der Beek, Y. Fasano, N.R. Cejas Bolecek, H. Pastoriza, D. Colson, and F. Rullier-Albenque, Phys. Rev. B **84**, 094517 (2011).

[47] C. Tarantini, S. Lee, Y. Zhang, J. Jiang, C.W. Bark, J.D. Weiss, A. Polyanskii, C.T. Nelson, H.W. Jang, C.M. Folkman, S.H. Baek, X.Q. Pan, A. Gurevich, E.E. Hellstrom, C.B. Eom, and D.C. Larbalestier, Appl. Phys. Lett. **96**, 142510 (2010).

[48] Y. Matsuda, T. Nagaoka, G. Suzuki, K. Kumagai, M. Suzuki, M. Machida, M. Sera, M. Hiroi, and N. Kobayashi, Phys. Rev. B **52**, R15749 (1995).

[49] T. Nagaoka, Y. Matsuda, H. Obara, A. Sawa, T. Terashima, I. Chong, M. Takano, and M. Suzuki, Phys. Rev. Lett. **80**, 3594 (1998).

[50] N.B. Kopnin, B.I. Ivlev, and V.A. Kalatsky, J. Low Temp. Phys. **90**, 1 (1993).

[51] A.T. Dorsey, Phys. Rev. B **46**, 8376 (1992).

[52] T. Nishizaki, T. Naito, and N. Kobayashi, Phys. Rev. B **58**, 11169 (1998).

[53] H.H. Wen, S.L. Li, Z.W. Zhao, H. Jin, Y.M. Ni, W.N. Kang, H.-J. Kim, E.-M. Choi, and S.-I. Lee, Phys. Rev. B **64**, 134505 (2001).

Table

Table I. Summary of the mixed-state Hall effect of a BaFe$_2$As$_2$:Co epitaxial film for *T* and *H* sweep modes.

| Sweep variable | Temperature | Magnetic field |
|---|---|---|
| Fixed parameter | Magnetic field | Temperature |
| Range | 1 → 9 T | 13 → 16 K |
| $\beta$ value | constant | monotonically increase |
| | 1.7–1.8 | 1.8 → 2.0 |
| Sign reversal | Not detected | |

Figure Captions

FIG. 1. (Color online) (a) Cross-sectional TEM image of BaFe$_2$As$_2$:Co epitaxial film. The Horizontal arrow indicates the heterointerface between the substrate and the BaFe$_2$As$_2$:Co film. The vertical arrows indicate the line defects. (b) EDXS line profile of chemical composition along the dashed horizontal line in FIG. (a) around a defect.

FIG. 2. (Color online) (a), (b) *T* sweep and (c), (d) *H* sweep results of $\rho_{xx}$ and $\rho_{xy}$ of a BaFe$_2$As$_2$:Co epitaxial film. Measurements were performed under fixed magnetic fields between 0 and 9 T (a), (b) and at fixed temperatures between 13 and 16 K (c), (d). Horizontal dotted lines show the measurement limits.

FIG. 3. (Color online) $T$ sweep scaling behavior between $\rho_{xy}$ and $\rho_{xx}$ under fixed magnetic fields varied from 1 to 9 T. Solid lines show fitted results using the relation $\rho_{xy}$ = A$\rho_{xx}^{\beta}$. Magnetic fields and the $\beta$ values are shown on top left.

FIG. 4. (Color online) $H$ sweep scaling behavior between $\rho_{xy}$ and $\rho_{xx}$ at fixed temperatures varied from 13 to 16 K. Solid lines are fitted results. Measurement temperatures and the $\beta$ values are shown on top left.

FIG. 5. (Color online) Vortex phase diagrams of (a) BaFe$_2$As$_2$:Co epitaxial film and (b) BaFe$_2$As$_2$:Co epitaxial film with YBCO[22,52] and MgB$_2$[53] normalized by $H_{c2}$ and $T_c$. Dotted and dashed lines indicate the irreversibility lines ($H_{irr}$) and $H_{c2}$, respectively, which are schematically shown in (c) as a general vortex diagram. (a) Vertical and horizontal lines show the results of the $T$ sweep (taken from Fig. 3) and the $H$ sweep measurements (taken from Fig. 4), respectively.

FIG. 6. (Color online) Vortex-penetration processes under various conditions for the $T$ sweep ($H$ is increased and then $T$ is increased) and $H$ sweep ($T$ is increased and then $H$ is increased) processes. $U$ denotes the effective pinning potential. $N_v$ and $N_p$ denote the numbers of penetrated vortices and strong pinning centers, respectively. Short and long vertical arrows denote pinned vortices and unpinned vortices, respectively. (a) Initial state at low $T$ (e.g., 5 K) and $H = 0$ ($N_v = 0$). (b) Final condition corresponds to a high $H$ (i.e., the final $N_v >> N_p$) and a relatively low $T$. For the $T$ sweep, the strong pinning centers are filled with vortices, and many extra vortices spill over upon applying high $H$

(upper left). Pinned vortices are maintained at higher $T$ but extra vortices are mobile (upper right). For the $H$ sweep, the effective pinning force weakens at higher $T$ (lower left); then vortices are introduced and most of the pinning centers are filled due to the condition $N_v >> N_p$, and the extra vortices are mobile (lower right). This final situation is the same as the $T$ sweep case. (c) Final condition corresponds to low $H$ (the final $N_v \sim N_p$) and high $T$. For the $T$ sweep, the penetrated vortices are trapped by the pinning centers upon applying $H$ (upper left). These trapped vortices are maintained at high $T$ (upper right), giving a strong pinning condition with $\beta < 2.0$. For the $H$ sweep, effective pinning force weakens at high $T$ (lower left); then vortices are introduced, but many vortices are not trapped and remain mobile (lower right), giving rise to a weak pinning condition with $\beta \sim 2.0$. The final state differs from the $T$ sweep case.

FIG. 7. (Color online) $H$ dependence of the Hall conductivity $\sigma_{xy} = \rho_{xy}/(\rho_{xy}^2 + \rho_{xx}^2)$ at 13–16 K. Inset shows $T$ dependence of $\sigma_{xy}$ under 1–9 T.

(a)

(b)

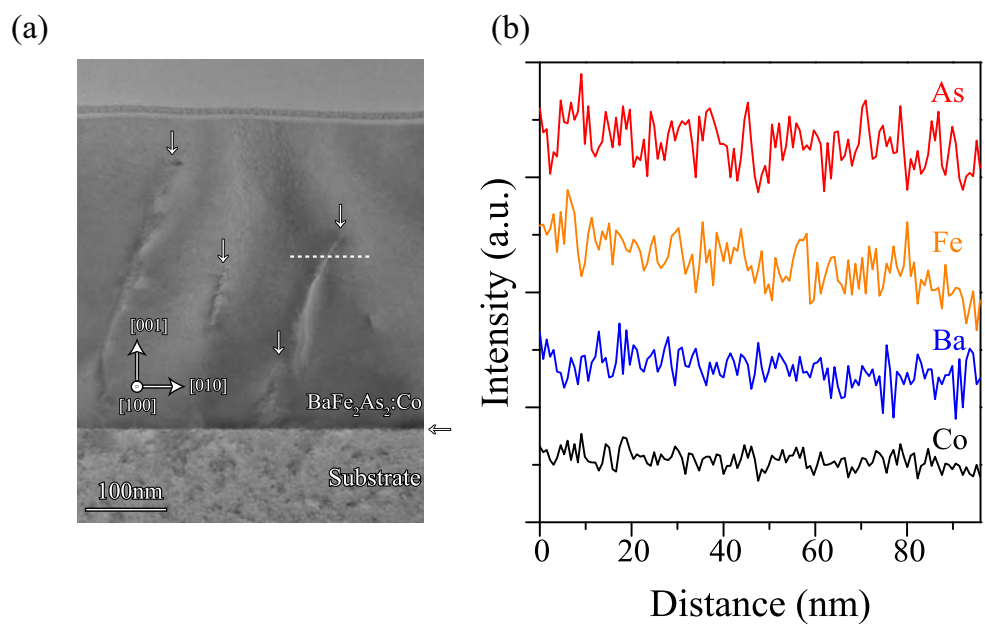

Fig. 1

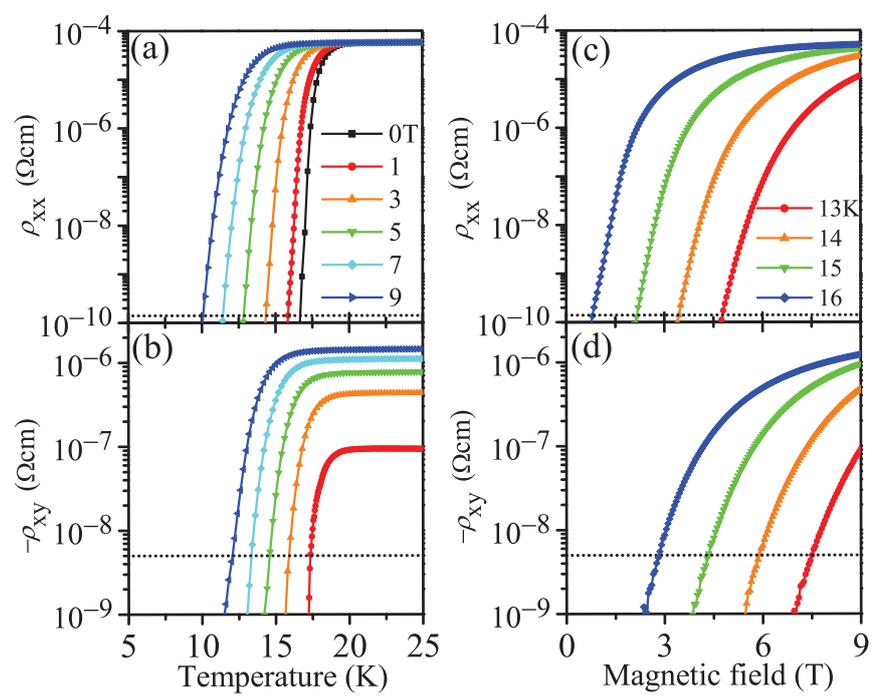

Fig. 2

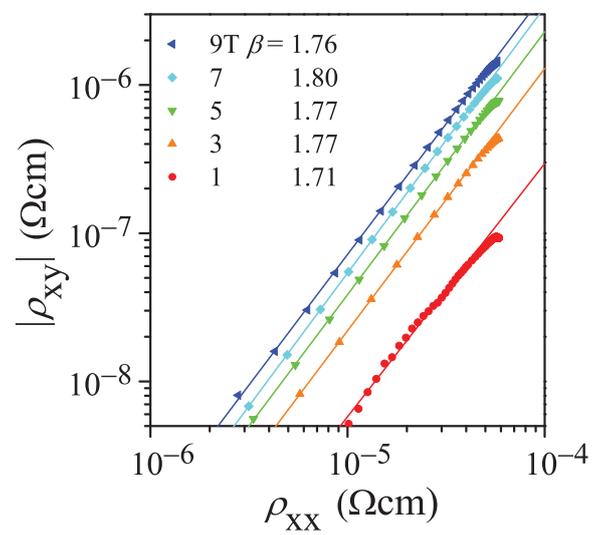

Fig. 3

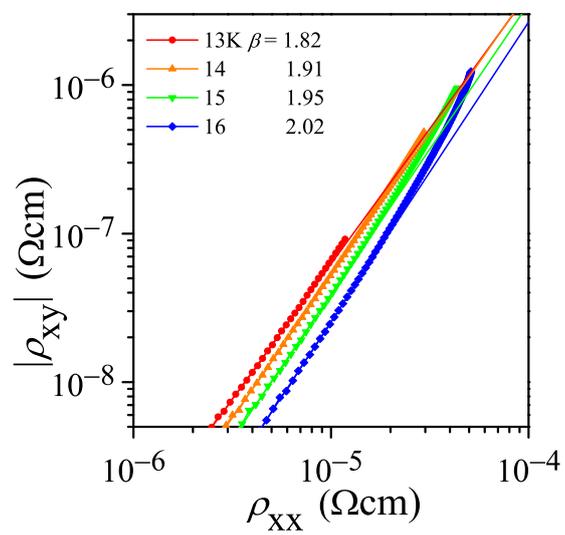

Fig. 4

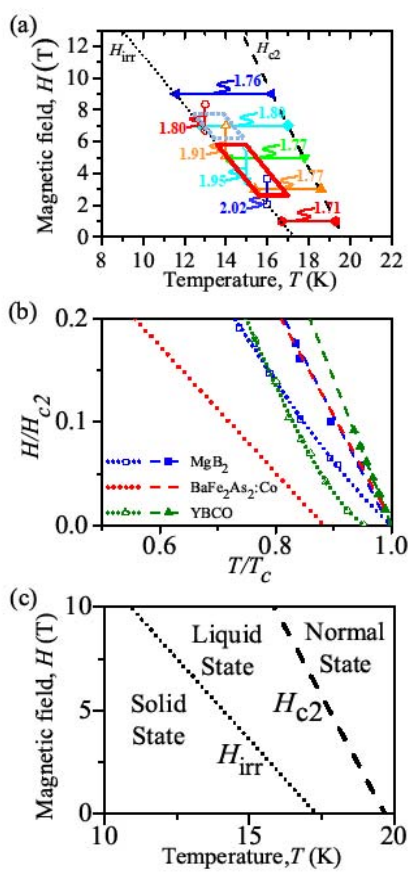

Fig. 5

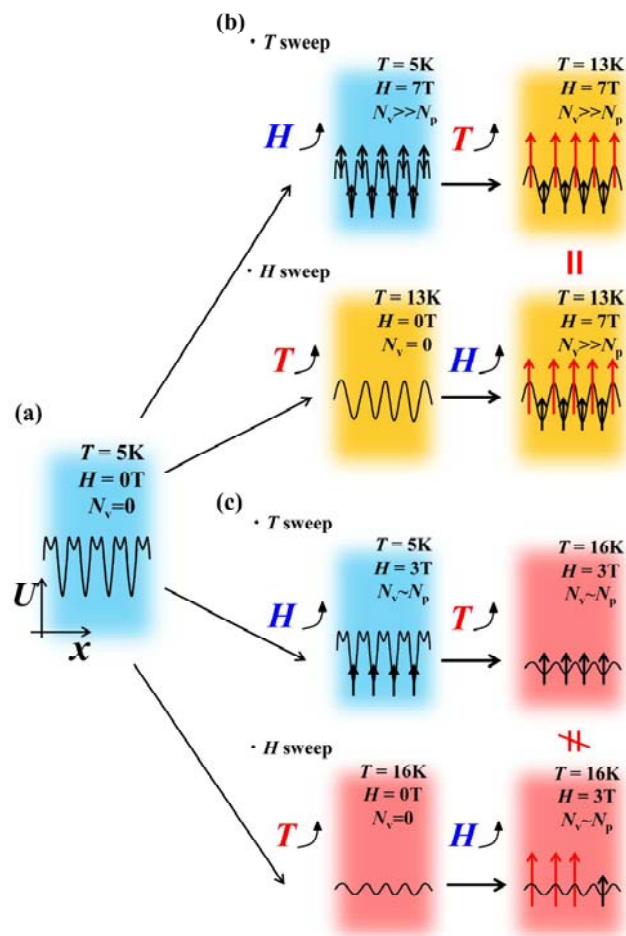

Fig. 6

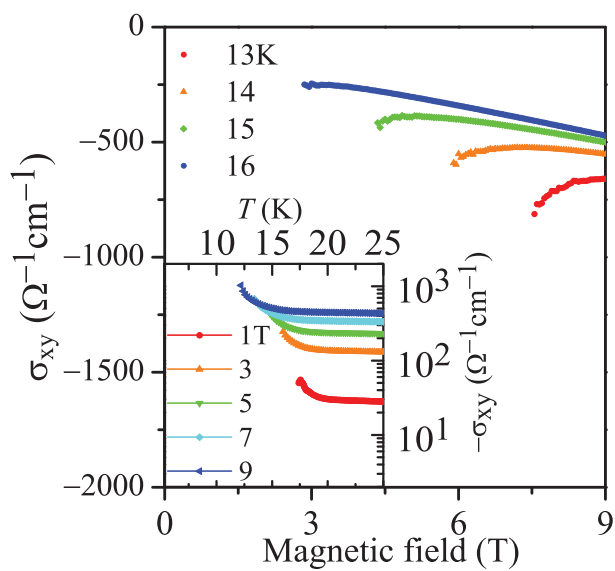

Fig. 7